\documentclass[aps,reprint,prl,superscriptaddress,showkeys,nofootinbib]{revtex4-1}
\usepackage{amsfonts, amsmath, amssymb, wasysym}
\usepackage[pdftex]{graphicx}

\usepackage{titlesec}

\titleformat{\section}
  {\large\bf\centering}{\thesection}{1em}{}
\titleformat{\subsection}
  {\bf\raggedright}{\thesubsection}{1em}{}
\titleformat{\subsubsection}
  {\bf\raggedright\it}{\thesubsection}{1em}{}

\newcommand{\AP}{\mathrm{AP}}
\newcommand{\eff}{\mathrm{eff}}
\newcommand{\acc}{\mathrm{acc}}
\newcommand{\raw}{\mathrm{raw}}

\newcommand{\noise}{\mathrm{noise}}
\newcommand{\recorded}{\mathrm{recorded}}
\newcommand{\expected}{\mathrm{expected}}

\begin{document}
\title{Only accessible information is useful: insights from gradient-mediated patterning}

\author{Mikhail Tikhonov}
\affiliation{Center of Mathematical Sciences and Applications, Harvard University, Cambridge, MA 02138, USA}
\author{Shawn C. Little}
\affiliation{Howard Hughes Medical Institute}
\affiliation{Department of Molecular Biology}
\author{Thomas Gregor}
\affiliation{Joseph Henry Laboratories of Physics}
\affiliation{Lewis-Sigler Institute for Integrative Genomics, Princeton University, Princeton, NJ 08544, USA}

\begin{abstract}
Information theory is gaining popularity as a tool to characterize performance of biological systems. However, information is commonly quantified without reference to whether or how a system could extract and use it; as a result, information-theoretic quantities are easily misinterpreted. Here we take the example of pattern-forming developmental systems which are commonly structured as cascades of sequential gene expression steps. Such a multi-tiered structure appears to constitute sub-optimal use of the positional information provided by the input morphogen because noise is added at each tier. However, the conventional theory fails to distinguish between the total information in a morphogen and information that can be usefully extracted and interpreted by downstream elements. We demonstrate that quantifying the information that is {\it accessible} to the system naturally explains the prevalence of multi-tiered network architectures as a consequence of the noise inherent to the control of gene expression. We support our argument with empirical observations from patterning along the major body axis of the fruit fly embryo. Our results exhibit the limitations of the standard information-theoretic characterization of biological signaling and illustrate how they can be resolved.
\end{abstract}
\maketitle




As an inspiring example of productive collaboration between computer science, physics and biology, information theory is gaining popularity as a tool to characterize performance of biological systems. Although is may not have become the ``general calculus for biology'', as predicted by Johnson in his 1970 review~\cite{johnson_70}, the scope of its applications has been steadily expanding: from the earliest work measuring the information content in DNA, RNA and proteins to topics like neuroscience, collective behavior, ecology, developmental biology, genetic regulation and signaling~\cite{waltermann_11,bowsher_14,levchenko_14,tkacikBialek_14}.

Specifically in the context of biochemical signaling, several recent reviews make compelling arguments that the mutual information between input and output of a signaling pathway is not just a useful quantity, but is in fact the ``only natural framework'' for characterizing the performance of such systems. However, implicit in these arguments is the assumption that the ``output'' in question is the final target of signaling, the functionally relevant phenotypic trait. Unfortunately, in biological applications of information theory information content is usually assessed for signals that constitute intermediate steps, most commonly transcription factors, for example, NF-$\kappa$B~\cite{cheong_11,selimkhanov_14} or Drosophila patterning cues~\cite{dubuis_13}. Such signals, however, still need to be interpreted by downstream processes. Therefore, the information they carry is useful only to the extent that it can be extracted and used by the system. As we will demonstrate, failure to recognize this can easily cause information-theoretic quantities to be misinterpreted.

To show this, we take the example of gradient-mediated patterning circuits. For a complex multicellular organism, the reliability of its developmental program directly determines the probability of reaching reproductive age; therefore, low error rate and/or high error tolerance are likely to be key determinants of the structures of developmental circuits~\cite{hironaka_2012,lander_2013}. Why, then, are so many patterning circuits structured as a cascade of several signaling steps, each of which is susceptible to loss of information due to noise inherent in biological control? We will see that treating information content of patterning cues as a one-size-fits-all method to characterize system performance erroneously predicts that a single-step readout strategy should be dominant in development. To understand the advantages of the multi-tiered architectures observed in real systems, it is essential to distinguish between the total information in a morphogen and information that can be usefully extracted and interpreted. We support our reasoning with experiments on the well-studied segmentation gene network responsible for anterior-posterior patterning in the \emph{Drosophila} embryo.

\begin{figure}[b!]
\centering
\includegraphics[width=0.95\linewidth]{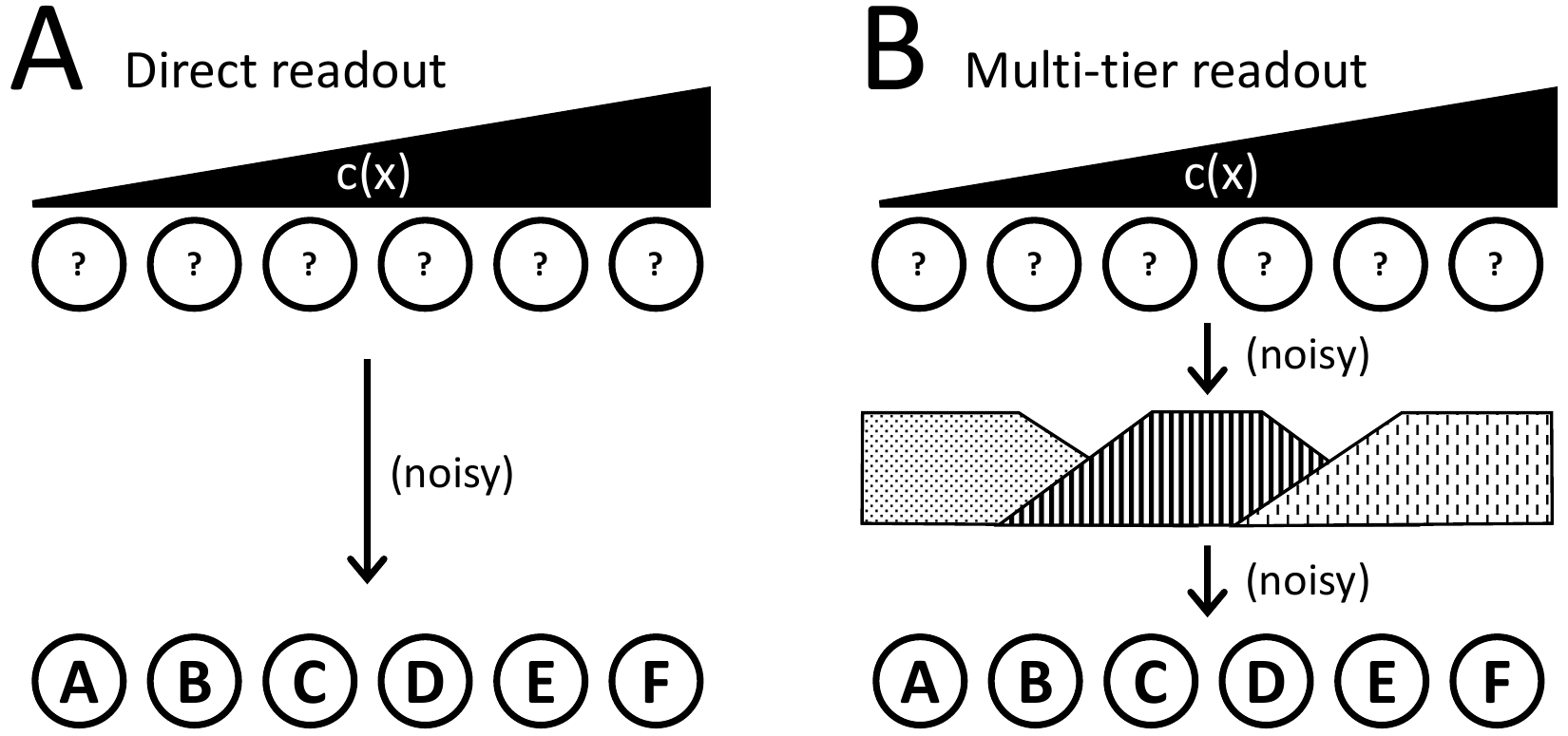}
\caption{Direct versus multi-tiered decoding strategies for gradient-mediated patterning. (A) Direct decoding: to reduce noise introduced by intrinsically variable gene expression, patterning proceeds through a single cycle of transcription and translation. Differences in morphogen input {\em c(x)} directly specify gene expression programs A-F along axis {\em x}. (B) Multi-tiered decoding: morphogen first elicits expression of short range diffusible factors in domains spanning several cells. These gene products then induce programs A-F through a second cycle of transcription/translation. The added step introduces additional gene expression noise, reducing patterning information compared to direct decoding (A).\label{fig:multitier}}
\end{figure}

\vspace{0.5cm}
\noindent\textbf{Multi-tier architecture in gradient-mediated patterning.}
In many developing embryonic systems, cellular identities are conferred by graded input signals that induce dose-dependent gene expression programs as outputs~\cite{rogers_2011,nahmad_2011}. Such graded inputs, termed morphogens, often function as diffusible molecules produced by a localized expression source~\cite{wartlick_2009,muller_2013}. Localized expression generates concentration gradients in a field of otherwise naive and identical cells (presented in simplified form as a one-dimensional array in Fig.~\ref{fig:multitier}). Cells activate specific expression programs in response to the local morphogen concentration $c(x)$. When $c$ correlates closely with distance $x$ from the source, such gradients carry a large amount of ``positional information''~\cite{wolpert_1969} quantified via the mutual information $I[c(x),x]$~\cite{shannon_48,dubuis_13}. In principle, a morphogen gradient carrying sufficient information could induce in each cell the gene expression program appropriate for its position, thus generating the required spatial arrangement of cell fates~\cite{gergen_86} (Fig.~\ref{fig:multitier}A). In the most straightforward model, assuming the input morphogen is sufficiently reproducible~\cite{gregor_2007}, local morphogen concentration is directly interpreted by each cell, i.e., the local input activates all genes required at a given position, with no additional cycles of gene expression modulation. A central tenet of information theory, the information processing inequality, states that each transmission or processing step can only reduce the total information contained in a signal. Direct decoding might therefore be expected to dominate in early development as the optimal strategy for transmitting positional information. This expectation seems all the more valid given the widespread observation that the processes of transcription and translation exhibit considerable intrinsic variability, or noise~\cite{munsky_2012,sanchez_2013}. Thus information loss in gene regulatory processes should be particularly notable.

Therefore, from the perspective of information theory, it is surprising that many gradient-based systems exhibit a multi-tiered architecture in which reiterated cycles of transcription and translation are required to attain patterning goals (illustrated in Fig.~\ref{fig:multitier}B). For example, in the vertebrate central nervous system, the unpatterned neuroectoderm exhibits a graded distribution of multiple diffusible signaling molecules. These signals subdivide the prospective brain into relatively large fore-, mid-, and hindbrain territories, which are then segmented into smaller subunits by additional signaling activity~\cite{lumsden_96,pera_14,raible_04}. Similar patterns of broad subdivision followed by short-range refinement are found during the specification of the vertebrate neural crest by reiterated rounds of extracellular signaling~\cite{patthey_14}; in the formation of segmented muscle precursors (somites) by FGF and Notch followed by short range Ephrin activity~\cite{saga_12,watanabe_10}; the dorsal-ventral patterning of the \emph{Drosophila} body axis, first by a gradient of NF-$\kappa$B activity (also called Dorsal) and then by members of the BMP family of secreted signaling molecules~\cite{little_06,rushlow_12}; and also in the fruit fly, the patterning of the anterior-posterior (AP) axis by gradients of diffusible transcription factors within the shared cytoplasm of the nuclear syncytium~\cite{gergen_86,driever_88,kornberg_93}.

These examples and others illustrate a common theme where long range signaling gradients subdivide a large field into smaller domains, within which the patterned expression of secondary factors establishes elaborated patterns (Fig.~\ref{fig:multitier}B). Since each cycle of transcription and translation introduces more noise, the widespread use of the multi-tiered architecture appears to conflict with the expectation that development should favor circuits exhibiting efficient information utilization.

This apparent conflict arises because Shannon's information content of a signal~\cite{shannon_48} has two important limitations. First, the information content of a patterning cue or other biological signal is defined locally in space and time, whereas its interpretation is non-local, and instead occurs over time and frequently involves diffusive signals. For this reason, the naive application of information processing inequality in these systems is incorrect, and the local, instantaneous information content in a signal does not in fact provide an upper bound for the performance of downstream processes interpreting this signal~\cite{tostevin_09,selimkhanov_14,tkacik_15}. Second, the same amount of information can be encoded in formats that are more or less easy for the system to access, since the interpreting circuit is itself subject to noise. Thus, the local information content of a signal is neither an upper bound nor a fair estimate of the amount of information this signal can ``transmit'' to the downstream circuit. This is well illustrated by the recent experimental work on ERK, calcium and NF-$\kappa$B pathways~\cite{selimkhanov_14}. If the output of any of these pathways is reduced to a single scalar, it is found to transmit very little information about the input. If the output is treated as a dynamical variable, its apparent information content increases considerably~\cite{tostevin_09}. Neither of these quantities, however, can be interpreted before it is established what fraction of that information can actually be extracted and used by the system. Here we use a simplified model to illustrate these limitations of what we call ``raw'' information content, contrasting it with ``accessible information'' that we introduce.


\section{Results}
\begin{figure}[t!]
\centering
 \includegraphics[width=\linewidth]{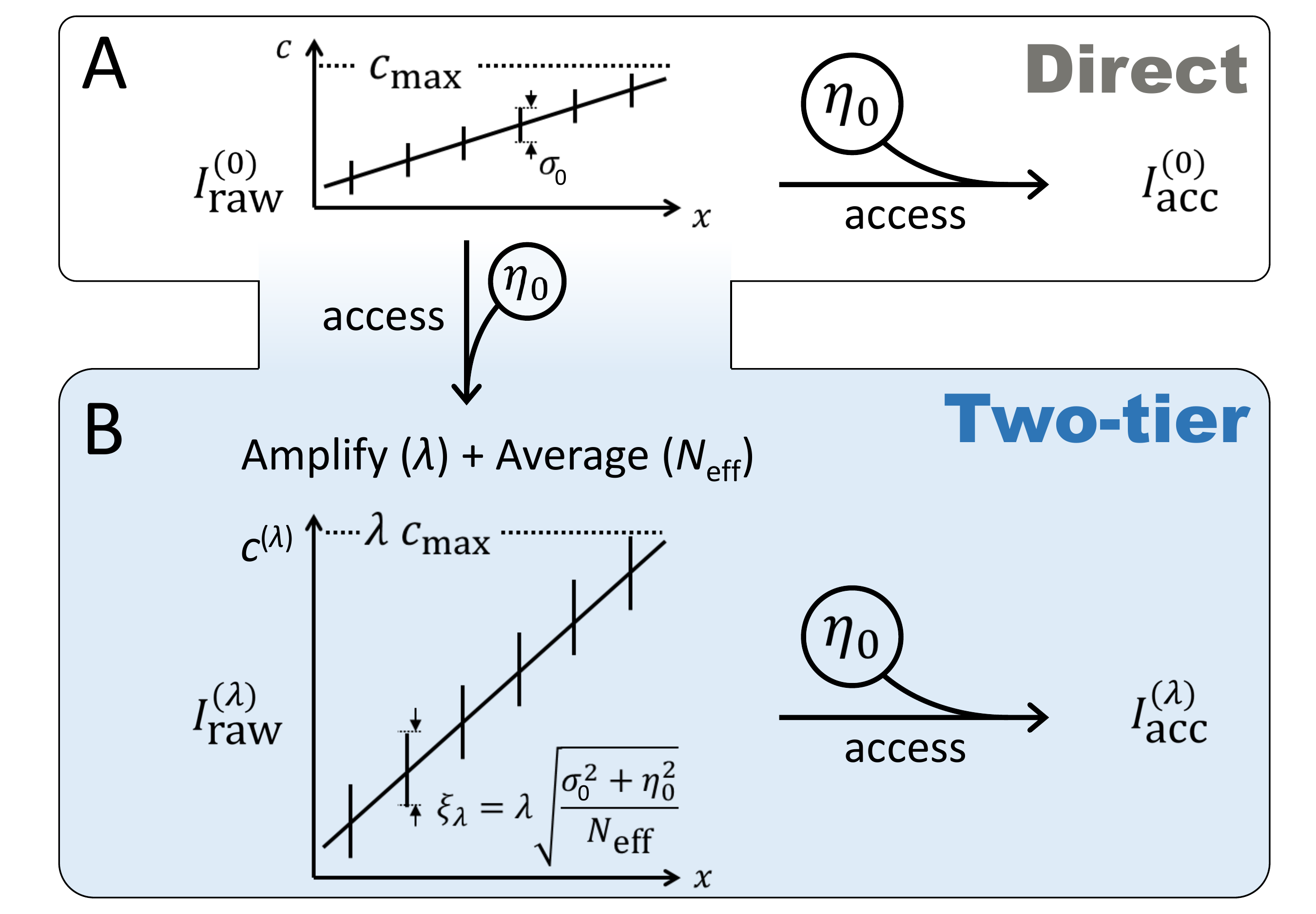}
\caption{\textbf{The two patterning strategies.} \textbf{A:} In the direct strategy, target genes are controlled directly by $c$. \textbf{B:} The two-tier strategy involves a second patterning factor $c^{(\lambda)}$; target genes are separated from the input by two tiers of ``access'' operations.  Left, raw information content. Right, accessible information content.\label{fig:argument}}
\end{figure}

\vspace{0.5cm}
\noindent\textbf{An abstract gradient response problem.} A one-dimensional array of cells $i$ located at positions ($0<x_i<L$) is exposed to a noisy linear gradient of an input morphogen $c(x)$ spanning the range $[0,c_\mathrm{max}]$. To build intuition, we will assume the noise of input $c(x)$ to be Gaussian, of constant magnitude $\sigma_0$, and uncorrelated between cells\footnote{The assumption of uncorrelated noise is intentionally strong. In a real system, correlated noise can be introduced, for example, by variations in the total amount of morphogen deposited maternally. These fluctuations, which cannot be reduced by averaging, lead to imperfect \emph{reproducibility} of morphogen activity at a given location across multiple embryos. Much work has focused on investigating the limitations imposed on patterning by this type of fluctuations~\cite{dubuis_13,tkacik_14,petkova_14}. In contrast, our model is applicable for understanding the effects of imperfect \emph{precision} of gene expression (at a given location within the same embryo). The distinction between ``raw'' and ``accessible'' information does not rely on the assumption of uncorrelated noise.}: $c(x_i)\equiv c_i = (x_i/L)\,c_\mathrm{max} + \sigma_i,$ where $\sigma_i$ are i.i.d., drawn from a Gaussian of width $\sigma_0$ (Fig.~\ref{fig:argument}A). Cells respond to morphogen $c(x)$ by modulating gene expression through intrinsically noise-prone signal transduction and transcription/translation processes. We will model this response as a composition of three steps, three elementary operations that constitute the ``toolkit'' with which cells can access and process information contained in patterning cues: \emph{access}, \emph{amplify}, and \emph{average}.

Let $g^{\mathrm{out}}$ be a gene product whose expression is controlled by $c(x)$. The simplest readout is achieved by placing gene $g^{\mathrm{out}}$ under the control of a promoter that is responsive to $c$ and by accumulating the output protein for some time $\tau$. In our model, we express the amount of $g^{\mathrm{out}}$ produced during this time by a cell $i$ as $g_i^\mathrm{out} = F(c^\mathrm{est})$, where $c^\mathrm{est}_i$ is a noisy estimate of the true concentration $c_i$ that the system could obtain in time $\tau$ (``access''), and $F$ is some deterministic input-output function (``amplify''); for simplicity, we first consider $F$ to be pure linear amplification with coefficient $\lambda$, denoted $F_\lambda$. The ``access'' operation is the key element of our framework. Specifically, we write
$$
c^\mathrm{est}_i=c_i+\eta_i,
$$
where $\eta_i$ reflects the intrinsic stochasticity of transcription and, in principle, many other noise sources. Here we will model $\eta_i$ simply as being drawn from a Gaussian distribution of width $\eta_0$. In other words, we postulate that each ``access'' operation takes time $\tau$ and comes at the price of corrupting the signal with extra noise of magnitude $\eta_0$.

The final toolkit operation is averaging. Because patterning systems typically act over durations that are long (hours) compared to the time required to synthesize mRNA and protein (minutes), cells can perform temporal averaging by allowing stable gene products to accumulate~\cite{little_13}: if $T$ is the time available for patterning, the system can effectively perform $T/\tau$ access operations. In addition, the production of soluble factors that can be shared between cells gives rise to spatial averaging~\cite{tenWolde_09,little_13}. Both types of averaging offer the system some capacity to perform multiple measurements of the input, which we capture formally by an averaging operator $G_{N_\eff}$. Here $N_\eff$ indicates the effective number of independent measurements, so that application of $G_{N_\eff}$ to a morphogen, by definition, reduces expression fluctuations by a factor $1/N_\eff$.

We distinguish between two patterning strategies. In the first (``direct strategy''; Fig.~\ref{fig:argument}A), cell-fate-specific target genes are controlled directly by $c$ and no other patterning factors are involved. Any available averaging mechanisms are applied to $c$ itself. In the second (``two-tier'') strategy, cells perform an amplifying readout of $c$ with input-output function $F_\lambda$ to establish a spatial profile of a second factor $c^{(\lambda)}$ (Fig.~\ref{fig:argument}B). The pattering time $T$ is spent on accumulating and averaging $c^{(\lambda)}$. Mathematically, in the two scenarios, the cell-fate-specific target genes are controlled by:
\begin{align}
  c^{(0)}&= G_{N_\eff}[c] &&\text{(direct strategy)}\label{eq:DirectStrategy}\\
  c^{(\lambda)}&= G_{N_\eff}[F_\lambda(c+\eta)]&&\text{(two-tier strategy)}\label{eq:TwoTierStrategy}
\end{align}
We now ask: when, if ever, does the noisy amplification step of the two-tier strategy provide a benefit to the system?

\newpage
\noindent\textbf{Standard information-theoretic considerations do not explain the benefits of amplification.}
The positional information carried by a linear morphogen $c(x)$ with dynamic range $c_\mathrm{max}$ and noise $\sigma_0$, which we call the ``raw information content'' of a gene expression profile, is given by
$$
I_\raw[c(x),x] = \ln\left(\frac{c_\mathrm{max}}{\sigma_0\sqrt{2\pi e}}\right)
$$
(see Supplementary Information). It depends only on the ratio $\phi=c_\mathrm{max}/\sigma_o$; for convenience, we define $\mathcal I(\phi) \equiv \ln\left(\frac{\phi}{\sqrt{2\pi e}}\right)$, which is an increasing function of $\phi$.

Let us compare the two patterning strategies from the point of view of the raw information content carried by the controlling signal. In the direct strategy~\eqref{eq:DirectStrategy}, the application of $G_{N_\eff}$ reduces the input noise to $\sigma_o/\sqrt{N_\eff}$ and so the controlling signal $c^{(0)}$ carries $I_\raw^{(0)}=\mathcal I\left(\frac{c_\mathrm{max}}{\sigma_0/\sqrt{N_\eff}}\right)$ bits of raw information. In the two-tier strategy~\eqref{eq:TwoTierStrategy}, the amplified profile $c^{(\lambda)}$ is characterized by noise
$
\xi_\lambda=\lambda \sqrt {\frac{\sigma_0^2+\eta_0^2}{N_\eff}},
$
and its raw information content is therefore
\begin{equation}\label{eq:amplifiedInfo}
I^{(\lambda)}_\raw=\mathcal I\left(\frac{\lambda c_\mathrm{max}}{\xi_\lambda}\right)=\mathcal I\left(c_\mathrm{max}\sqrt{\frac{N_\eff}{\sigma_0^2+\eta_0^2}}\right) < I_\raw^{(0)}.
\end{equation}

Averaging mitigates the loss of positional information when using a noisy readout~\cite{tenWolde_09}. If $N_\eff$ is sufficiently large, the amplified and averaged profile carries even more information than the original input. (Note that the information processing inequality is not violated, as it states only that the output cannot carry more information than $N_\eff$ independent copies of the input.) Nevertheless, applying averaging directly to the input (the direct strategy) always yields more raw information; thus, the multi-step scenario appears inferior to a direct readout.

In real systems, the three operations we treat as independent may be mechanistically linked. For example, if $c(x)$ is an intracellular factor while spatial averaging requires a small diffusible molecule, then performing an extra readout can provide access to an otherwise unavailable averaging mechanism. By assuming that the two strategies~\eqref{eq:DirectStrategy} and~\eqref{eq:TwoTierStrategy} can benefit from equal amounts of averaging, which in our model simply reduces expression noise and is obviously beneficial, we can focus specifically on the effect of signal amplification. Multi-tier patterning proceeds through rounds of amplification: small differences in input result in large differences in gene expression so as to establish increasingly sharp boundaries delimiting expression domains~\cite{gilbert}, yet in our expression~\eqref{eq:amplifiedInfo} for the information content of the amplified profile $c^{(\lambda)}$, the amplification factor $\lambda$ cancels out. Thus, considerations based on raw information content fail to explain the prevalence of signal amplification.

\vspace{0.5cm}
\noindent\textbf{The benefits of the multi-tiered strategy lie in making the ``raw'' information more accessible.}
The benefits of amplification and the advantages of the multi-tier strategy become clear when we observe that, due to the intrinsic noise in the regulatory readout, the raw information content is an inadequate measure of a morphogen's usefulness to the system. The purpose of a morphogen is to activate downstream processes; the relevant quantity is therefore not the amount of information a morphogen carries, but the amount of information it can transmit to its downstream targets. Since biological control is intrinsically noisy, the two quantities are distinct.

\begin{figure}[t!]
\centering
\includegraphics[width=\linewidth]{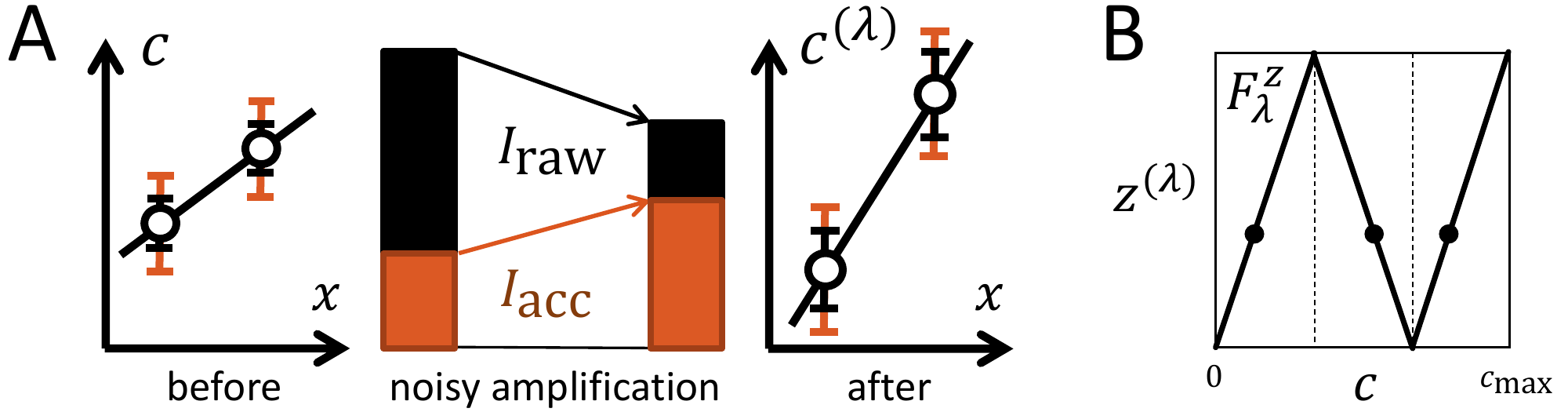}
\caption{\textbf{A:} Noisy amplification can increase accessible information even if raw information is reduced. Inner error bars are the signal variability and increase when amplification adds new noise, reducing $I_\raw$. Outer error bars represent the signal observed by the noisy cell machinery (corrupted by noise $\eta_0$). After amplification, the relative importance of $\eta_0$ is reduced, increasing $I_\acc$. \textbf{B:} The ``segmentation'' input-output function $F^z_\lambda$ for integer $\lambda$ (here $\lambda=3$) preserves the dynamic range of morphogen concentration. Locations such as those indicated by dots now have identical expression levels of $z^{(\lambda)}$ (the $y$ axis), but can be distinguished using the input morphogen $c$ (the $x$ axis on this plot).\label{fig:cartoon}}
\end{figure}

Our model was designed to make this particularly clear: since the system can never access the true concentration $c$, but only a noisy estimate $c^{\mathrm{est}}$, $I_\raw[c]$ is beyond the system's reach. We define \emph{accessible} information in a morphogen $I_\acc$ as the amount of information the system can access in time $\tau$:
\begin{equation}\label{eq:defAcc}
I_\acc[c]\equiv I_\raw[c^{\mathrm{est}}] = I_\raw[c+\eta],
\end{equation}
where $\eta$, again, is a Gaussian noise of magnitude $\eta_0$ within our model.

The amount of accessible information provided by the direct strategy (Fig.~\ref{fig:argument}B) is given by
\begin{equation}\label{eq:accNull}
I_\acc^{(0)}= \mathcal I\left(\frac{c_\mathrm{max}}{\sqrt{\frac{\sigma_0^2}{N_\eff}+\eta_0^2}}\right)
\end{equation}
whereas for the amplified profile $c^{(\lambda)}$ it is
\begin{equation}\label{eq:accAmpl}
I_\acc^{(\lambda)}=\mathcal I\left(\frac{\lambda c_\mathrm{max}}{\sqrt {\lambda^2 \frac{\sigma_0^2+\eta_0^2}{N_\eff}+\eta_0^2}}\right)
=\mathcal I \left(\frac{c_\mathrm{max}}{\sqrt {\frac{\sigma_0^2+\eta_0^2}{N_\eff}+\frac{\eta_0^2}{\lambda^2}}}\right).
\end{equation}
The amplification factor $\lambda$ no longer cancels out in ~\eqref{eq:accAmpl}; amplifying dynamic range is beneficial, since it reduces the relative importance of the intrinsic readout noise (Fig.~\ref{fig:cartoon}A). Comparing~\eqref{eq:accNull} and~\eqref{eq:accAmpl}, we find that the extra tier of noisy amplification is beneficial if and only if
\begin{equation}\label{eq:condition}
\eta_0^2\left(1-\frac 1{N_\eff}-\frac 1{\lambda^2}\right)>0
\end{equation}
Note that the condition~\eqref{eq:condition} is never satisfied if $N_\eff=1$ (no averaging) or $\lambda=1$ (no amplification). Intuitively, our argument demonstrates that the patterning system is a mechanism that invests some effort into making a careful measurement ($N_\eff>1$) and encodes this information in a more accessible format where steeper concentration changes ($\lambda>1$) can be interpreted with a faster, and therefore noisier readout. This mechanism is useful precisely because regulatory readout is intrinsically noisy, otherwise direct readout would have been the better strategy. In other words, to understand the purpose of the patterning system, it is essential to distinguish between the total information in a morphogen and information that can be usefully extracted and interpreted.

\vspace{0.5cm}
\noindent\textbf{Multiple tiers improve gradient interpretation even when raw information decreases.}
So far we considered the information content (raw or accessible) in each tier separately. However, in principle, downstream processes could access all patterning cues and not simply the final tier~\cite{tierIntegrate1,tierIntegrate2}. As a result, extra readout tiers can be beneficial even when they carry very little information on their own.

To see this, consider the input-output function $F^z_\lambda$ depicted in Fig.~\ref{fig:cartoon}B. In some respects, it is more realistic than the purely amplifying linear readout $F_\lambda$ considered above, since real patterning systems must operate within a limited global dynamic range of morphogen concentrations. Let $z^{(\lambda)}$ be the morphogen profile established by the new $F^z_\lambda$-shaped readout of $c$; it has noise magnitude $\xi_\lambda$ (same as the noise in $c^{(\lambda)}$), but is folded onto itself $\lambda$ times, reminiscent of the spatially reiterated expression of genes involved in \emph{Drosophila} axis segmentation. Repeatedly using the same output values at multiple positions naturally reduces mutual information between the output concentration and position:
$$
\begin{aligned}
I_\raw[z^{(\lambda)}]&=I_\raw[c^{(\lambda)}]-\ln\lambda\\
I_\acc[z^{(\lambda)}]&=I_\acc[c^{(\lambda)}]-\ln\lambda.
\end{aligned}
$$

However, the $\lambda$ locations with identical concentrations of $z^{(\lambda)}$ are made distinguishable by the original morphogen~$c$ (Fig.~\ref{fig:cartoon}B). Therefore, the \emph{joint} information that the original and the amplified profiles together provide about a cell's location is the same for $F^z_\lambda$ as it was for $F_\lambda$:
$$
I\big[\{c,z^{(\lambda)}\}, x\big]=I\big[\{c,c^{(\lambda)}\}, x\big]
$$
Replacing information content of a single profile by this joint information, our argument demonstrating that amplification increases accessible information can now be repeated verbatim~\cite{noteJointAcc}, and we again find that the extra readout is beneficial as long as~\eqref{eq:condition} is satisfied. Note, however, that on its own, $z^{(\lambda)}$ may carry \emph{less} information than the original morphogen $c$. The easiest way to see this is to compare their noise levels:
$$
\left(\frac{\xi_\lambda}{\sigma_0}\right)^2 = \frac{\lambda^2}{N_\eff}\left(1+\frac{\eta_0^2}{\sigma_0^2}\right)
$$
If the effect of amplification is stronger than that of averaging, we find $\xi_\lambda/\sigma_0>1$. In this scenario, the amplified profile $z^{(\lambda)}$ has the same dynamic range but lower precision than the original morphogen $c$, and therefore, on its own, carries less information (whether raw or accessible). This shows that evaluating the usefulness of a particular cue from information-theoretic standpoint can lead to misleading results, unless all other relevant cues (which are often hard to establish) are taken into account simultaneously. Here, we demonstrated that systems can benefit from multi-tiered interpretation even in cases where intermediate steps occur at a net loss of information, increasing noise.

\begin{figure*}[t!]
\centering
 \includegraphics[width=\textwidth]{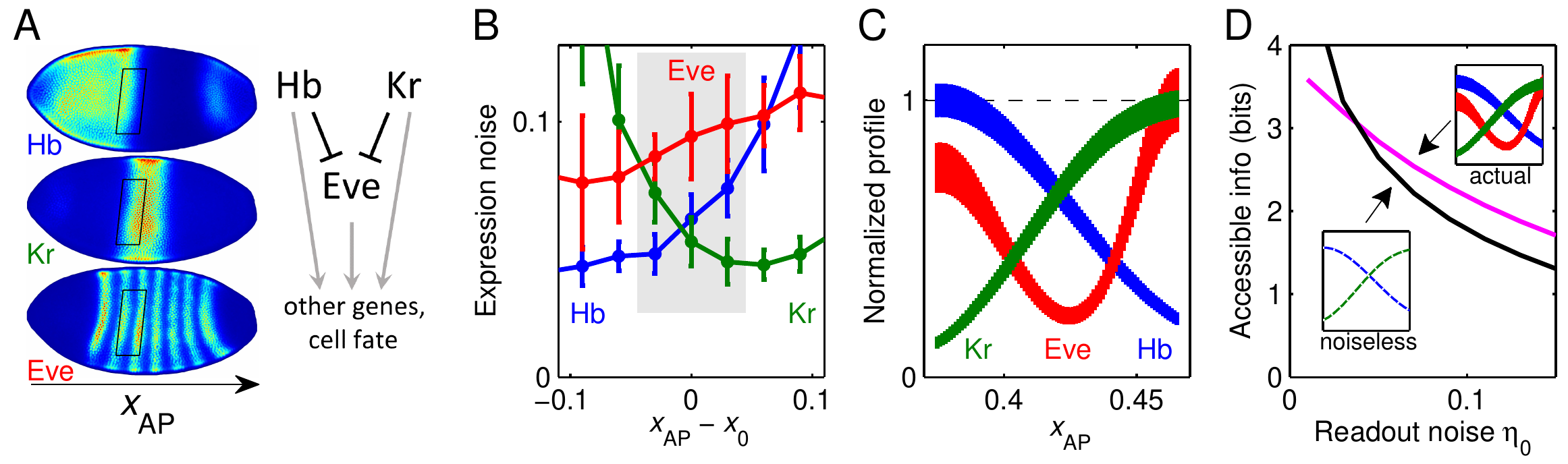}
\caption{\textbf{A:} Immunostaining of three antero-posterior (AP) axis patterning genes in the same embryo. Rather than specifying cell fate directly, the ``gap genes'' such as \emph{hunchback} (Hb; top) and \emph{Kr\"uppel} (Kr; middle) control ``pair-rule'' genes such as \emph{even-skipped} (Eve, bottom). Both tiers regulate other genes further downstream. Boxes indicate the selected region of interest (ROI), where at this time, Hb and Kr are the only relevant inputs to Eve, as shown on the cartoon. \textbf{B:} Within the ROI (shaded), Eve exhibits higher expression noise than either Hb or Kr. Expression noise computed as RMS difference between expression level of a nucleus and its immediate dorsal or ventral neighbor (see Methods), plotted against AP distance from the Hb/Kr boundary (denoted $x_0$). Error bars are standard deviation over $N=8$ embryos. \textbf{C:} Idealized morphogen profiles, restricted to the ROI. Profile shape obtained as smooth spline-fit to expression values and noise magnitudes calculated for the profiles of panel A after projection onto the AP axis. \textbf{D:} For all but the lowest readout noise magnitude, joint accessible information content in the triplet (Hb,Kr,Eve) exceeds the accessible information provided by Hb and Kr alone, even in an extreme hypothetical case when they are rendered entirely noiseless.\label{fig:data}}
\end{figure*}

\vspace{0.5cm}
\noindent\textbf{The multi-tier structure of \textit{Drosophila} segment patterning increases information accessibility.}
In this system, segmentation of the AP axis proceeds through four tiers of gene activity, termed maternal gradients, gap genes, pair-rule genes, and segment polarity genes~\cite{kornberg_93}. The sequential activity of each tier subdivides the naive blastoderm into smaller domains of gene expression with increasingly sharp boundaries, culminating in the designation of each row of cells with its own unique set of expressed genes (Fig.~\ref{fig:data}A). This process is subject to transcriptional noise with a large intrinsic component~\cite{little_13}, as well as several other noise sources with different signatures~\cite{tkacik_08,krivega_2012,kwak_2013,maheshri_2007}. No single value of $\eta_0$ adequately characterizes such readout noise. Nevertheless, we can gain important insight by computing $I_\acc^{\eta_0}[c]$ as a function of $\eta_0$, treating it as a variable parameter: the decay of $I_\acc^{\eta_0}[c]$ with $\eta_0$ characterizes the tolerance to added noise of the information encoded in the morphogen (or set of morphogens) $c$. Applied to gene expression data from the early \emph{Drosophila} segmentation gene network, this analysis will show how our simple model explains the use of multi-tier gradient interpretation in a real system (Fig. 4).

We focus on a particular node in this network whereby, in early embryos, two gap genes, \textit{hb} and \textit{Kr}, regulate a pair-rule gene \textit{eve}. For $0.37<x_\AP<0.47$, where \textit{Kr} and \textit{hb} expression form opposing boundaries, they are jointly responsible for creating the trough between \textit{eve} stripes 2 and 3; other inputs to \textit{eve} are negligible in this region at this time~\cite{kraut_91,small_96}. Protein levels are measured simultaneously in each nucleus by a triple immunostaining experiment (Fig.~\ref{fig:data}A) in $N=8$ single embryos. We determine the expression noise of each gene by comparing levels in a given nucleus with those of its immediate dorsal and ventral neighbors (see Methods).

In the defined region of interest, \textit{eve} expression noise is higher than the respective noise in \textit{hb} or \textit{Kr} expression (Fig.~\ref{fig:data}B). The information content of \textit{eve} must therefore be lower than that carried by either of its two inputs. Due to the curvature of the embryo  (Fig.~\ref{fig:data}A), the positional information of a real morphogen is only approximately related to that derived from projection onto the imaginary AP axis. Therefore, to estimate the information content for each of the three genes, we consider ``idealized'' Gaussian-noise profiles (panel C) with mean and noise obtained by smoothing the measured values in real embryos. The idealized profiles are normalized to the same maximum and are, by construction, functions of $x_\AP$ carrying positional information $I(c(x_\AP), x_\AP)$. Restricted to the region of interest, the information content of Hb and Kr is respectively 2.6 and 2.7 bits, whereas the larger noise of Eve reduces its information content to only 2.0 bits. Why, then, does the system use Eve to regulate downstream processes, rather than utilizing Kr and Hb directly?

The answer becomes clear when we consider the accessibility of information encoded in these morphogens, namely $I_\acc^{\eta_0}$ as a function of $\eta_0$ (panel D). A patterning strategy lacking Eve can access only Hb and Kr. Even if some hypothetical filtering mechanism could reduce their expression noise to arbitrarily low level, the readout noise magnitude $\eta_0>0$ imposes an upper bound that $I_\acc^{\eta_0}[c_\mathrm{Hb}, c_\mathrm{Kr}]$ must satisfy. This corresponds to the information in a hypothetical pair of noiseless Hb and Kr and cannot be achieved in practice; it is a theoretical best-case scenario for any strategy lacking Eve.

When the readout noise $\eta_0$ is zero, $I_\acc^{\eta_0}$ coincides with the raw information content, which for perfectly noiseless Hb and Kr would be infinite. However, as readout noise increases, the performance bound becomes finite and drops quickly (black curve). This behavior contrasts with the joint accessible information of the triplet (Hb,Kr,Eve) (magenta) as calculated using the actual measured noise of each of the three profiles. The accessible information content in the triplet is, of course, always finite, but it is also more tolerant to readout noise: due to the steeper slopes of the Eve profile, as $\eta_0$ increases, the accessible information content of the triplet (Hb,Kr,Eve) decreases slowly; importantly, more slowly than the black curve. Therefore, a crossing point is observed, whose presence does not qualitatively depend on the specifics of the readout noise model (e.g.\ absolute noise magnitude can be replaced by fractional). Remarkably, although Eve is measurably noisier than either of its inputs, its presence enables the system to access more information than could have been extracted from Hb and Kr alone, even if these inputs could be rendered perfectly noiseless. In practice, the enhancers of the pair-rule genes also contain binding sites for maternal transcription factors~\cite{tierIntegrate1,tierIntegrate2}, which may lead to a further increase in the precision of gene expression. However, our framework demonstrates that even if Eve were regulated by Hb and Kr only, and so were fully redundant in the standard information-theoretic sense, the additional tier would still confer an advantage, because transcription is intrinsically noisy.


\section{Discussion}

The \emph{Drosophila} patterning network has been described as performing a ``transition from analog to digital specification'' of cell identity~\cite{gilbert}. The ``digital'' metaphor has its limitations: even for Eve, the graded distribution within gene expression domains contains information~\cite{dubuis_13}; nevertheless, it expresses the correct intuition that the final pattern is more tolerant to noise. Importantly, the standard information-theoretic formalism does not capture this intuition: for instance, the profile depicted in Fig.~\ref{fig:cartoon}B has the \emph{same} information content for all $\lambda$. Noise tolerance --- a critically important feature in biological systems --- becomes manifest only when the readout process is considered explicitly, for example, as we have done in our definition of accessible information. This point is implicit in the theoretical work investigating the so-called ``input noise''~\cite{tkacik_08}, but has not been emphasized. This is because in a theoretical discussion of an abstract biochemical circuit, the quantities for which information is computed are easily postulated to be the complete input and the final output; in this manner, valid theoretical results can be derived without a concern for information accessibility (for some recent examples, see~\cite{bowsher_13,deronde_14}). However, when information-theoretic arguments are applied to experimental data where the measured quantity is only an intermediate step, e.g. a transcription factor regulating downstream events, the question of information accessibility can no longer be neglected.

For example, it has been suggested that certain signaling circuits may have evolved towards optimal information transmission~\cite{tkacikBialek_14,levchenko_14}. Although the argument is plausible, applying it in practice requires caution. Consider, once again, the example of a developmental circuit. If the entire set of functional (cell-fate specific) genes were to be included into consideration, then information transmission from the input to this entire layer of functional genes would be a plausible objective function for this whole network to maximize, under some ``bounded complexity'' constraint penalizing solutions where hundreds of cell-fate specific genes are all controlled by highly complex enhancers with combinatorial, cooperative regulation. However, the usual, more economical approach does not consider the full set of hundreds of cell-fate determining genes. Instead, it recognizes that the bulk of the patterning task is accomplished by a small subset of dedicated genes that engage in complex cross-regulation to establish the pattern that all other genes can then interpret simply. If we focus only on this core subset, the ``economy of complexity'' constraint is conveniently imposed by construction. We must realize, however, that maximizing information transmission to the target genes (downstream of the patterning core) imposes a different requirement onto this core circuit than merely efficient information transfer within the core itself. Instead, the core circuit must function as a format converter, re-encoding information at its input into a format that can be accessed with a simpler and faster readout, that of a patterning cue by a functional gene.

Curiously, it has been shown that in small networks with a realistic model of noise, maximizing raw information transmission leads to network structures exhibiting features such as tiling of patterned range with amplifying input/output readouts~\cite{optimFlow1,optimFlow2,optimFlow3}, i.e.\ features that tend to also make information more accessible, even though the optimization scheme employed in these studies did not specifically consider the encoding format. This remarkable coincidence, however, should not obscure the fact that ultimately the two tasks --- maximizing information transmission and re-encoding it in a more accessible format --- could be conflicting.

Information theory is a powerful tool; its formalism does not, however, aim to replace considerations of what constitutes useful information or how it might be used by the system. As it is gaining popularity in biological applications, it is important to remember that for a channel $X\mapsto Y$, the relation between mutual information $I(X,Y)$ and the ability to use $Y$ to determine $X$ is only asymptotic: Shannon~\cite{shannon_48} proved that it is the maximum rate of error-free communication via this channel, \emph{in the limit of infinite uses} of the channel. Importantly, in development and biological signaling, the number of channel uses (e.g. integration time of the signal) is fundamentally finite~\cite{bowsher_14}. Further, Shannon's results assumed an encoder/decoder of infinite computational power~\cite{shannon_48}. This asymptotic rate is never in fact achieved in practice~\cite{macKay}, but in biological context, performance is constrained even further, since the ``encoding scheme'' is usually limited to measuring the same signal multiple times. In communication theory, this bears the name of  ``repetition code'' and is formally classified as a ``bad code'', i.e.\ a code that does not attain Shannon's bound even asymptotically. This means that extracting all the ``raw'' information from a signal is impossible even in principle. For example, a signaling pathway with capacity of 1 bit is never sufficient to make a reliable binary decision~\cite{bowsher_14}, and therefore should not be conceptualized as a binary switch.

As illustrated here, making the distinction between ``raw'' and ``accessible'' information will be crucial for understanding the architecture and function of patterning and signaling circuits. More work is required: our definition of accessible information relied on a simplistic noise model; in general, quantifying the usefulness of information-bearing signals in contexts where channel uses are limited will require reinstating considerations of rate/fidelity tradeoff, which Shannon could eliminate by taking the limit of infinite-time communication. Nevertheless, information theory remains a most adequate framework to address these issues, provided it is extended to quantify both the amount and accessibility of information. Our work provides a step in this direction and demonstrates how the extended framework naturally explains a global architectural property shared by diverse patterning circuits.

\begin{acknowledgments}
We thank Ariel Amir, William Bialek, Michael Brenner, Chase Broedersz, Ted Cox, Paul Francois, Anders Hansen, Ben Machta, Gasper Tkacik, Eric Wieschaus and Ned Wingreen for helpful discussions and comments on the manuscript. This work was supported by NIH grants P50 GM071508 and R01 GM097275, NSF grants PHY-0957573, PHY-1305525, and Harvard Center of Mathematical Sciences and Applications.
\end{acknowledgments}

\appendix
\cleardoublepage
\onecolumngrid
\section{Supplementary Information}
 \renewcommand{\tabcolsep}{0.5cm}
 \setcounter{figure}{0}
 \renewcommand{\thefigure}{S\arabic{figure}}
 \renewcommand{\thetable}{S\arabic{table}}
\twocolumngrid
\subsection{Information carried by a linear morphogen gradient}
For a linear morphogen $c(x)$ spanning the range $[0,c_\mathrm{max}]$, with constant Gaussian noise $\sigma_0$, the information content is given by
$$
I_\raw[c]\equiv I[c(x),x] = \ln\left(\frac{c_\mathrm{max}}{\sigma_0\sqrt{2\pi e}}\right).
$$

To show this, we apply the definition of the mutual information:
$$
I[c(x),x] = H[P_c] - H[P_{c|x}]
$$
Here $P_c$ is the probability distribution of $c$ (which is uniform between 0 and $c_\mathrm{max}$); $P_{c|x}$ is the conditional distribution of the concentration of $c$ given $x$ (which is Gaussian of width $\sigma_0$), and $H[P]$ is the differential entropy of a probability distribution $P$:
$$
H[P]\equiv-\int P(z)\ln P(z)\,dz=-\langle \ln P\rangle_P.
$$

Clearly, $H[P_c]=\ln c_\mathrm{max}$. The second term is the entropy of a Gaussian distribution $P_{\sigma_0}$ of width $\sigma_0$:
$$
P_{\sigma_0}(z)=\frac 1{\sqrt{2\pi\sigma_0^2}}\exp\left(-\frac{z^2}{2\sigma_0^2}\right)
$$
and therefore:
\begin{multline}
    H[P_{c|x}]=-\langle \ln P_{\sigma_0}(z)\rangle_z
    = \ln \sqrt{2\pi\sigma_0^2}+\left\langle\frac {z^2}{2\sigma_0^2} \right\rangle_z\\
    = \ln \sqrt{2\pi\sigma_0^2}+\frac 12 = \ln\left(\sigma_0\sqrt{2\pi e}\right).
\end{multline}
Putting this together, we find:
$$
I[c(x),x] = H[P_c] - H[P_{c|x}]=\ln\left(\frac{c_\mathrm{max}}{\sigma_0\sqrt{2\pi e}}\right).
$$

\subsection{Experimental procedures}

Antibody staining was performed using procedures and antisera described in~\cite{Dubuis:2013} and~\cite{DubuisTkacik:2013}. Confocal microscopy was performed at 12 bit resolution on a Leica SP5 with a 20x HC PL APO NA 0.7 immersion objective at 1.4x magnified zoom using pixels of size 135 x 135 nm covering an area of 554x554 mm. For each embryo, 17 images slices were obtained at a $z$ interval of 4 microns, spanning approximately 50\% of embryo thickness. All data were collected in a single acquisition cycle using identical scanning parameters.

\subsection{Estimating expression magnitude (image processing)}
\begin{figure}[b!]
\centering
\includegraphics[width=\linewidth]{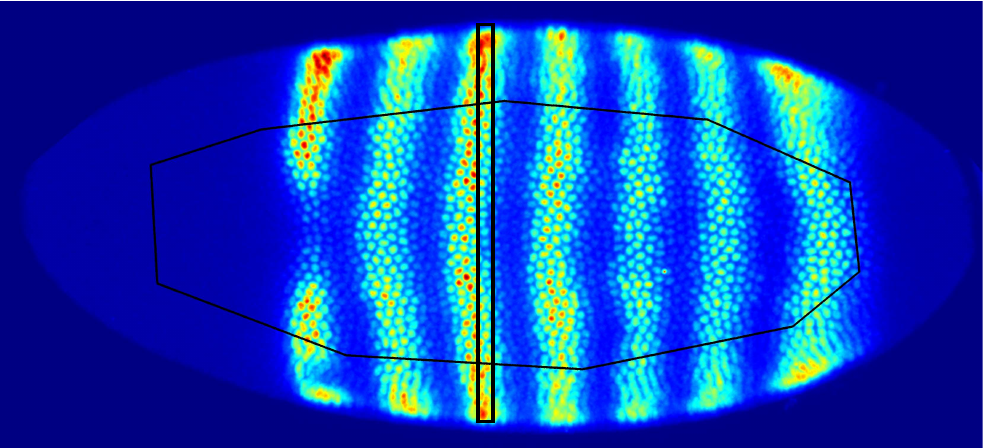}
\caption{\small{Example of projected image (Eve). Black polygon indicates the analysis region, manually selected to exclude distorted areas close to the embryo edge. Rectangle indicates nuclei with the same projected coordinate onto the AP axis. Even in this perfectly ventral view of the embryo that minimizes the effects of stripe curvature (compare with Fig. 4A in the main text), the expression stripes are not exactly perpendicular to this axis.}\label{fig:embryo8}}
\end{figure}

The immunostaining procedure described above yields confocal stacks of images where pixel intensity corresponds to the recorded fluorescence level. Stacks were converted into projected Hb, Kr and Eve images (such as displayed on Fig.~4A) as the maximum projection of Gaussian-smoothed frames. The width of the averaging kernel (8 pixels, corresponding to approximately 1 $\mu$m) was smaller than the radius of the nuclei, therefore for pixels close to the nucleus center the averaging volume was wholly within the nucleus. Smoothing frames prior to maximum projection ensured robustness against imaging noise.

In each of $N=8$ embryos, the location of nuclei was identified manually. For each of the projected images (Hb, Kr and Eve), we recorded the highest intensity value within 5 pixels of nuclei center locations as the fluorescence intensity in that nucleus. Allowing for a 5-pixel ``wiggle room'' ensured robustness against registration errors across color channels, as well as against errors in the manual selection of nuclei center locations. The recorded intensity values were corrected for background autofluorescence by subtracting the mean intensity recorded in nuclei located in non-expressing regions of the embryo. The background-corrected fluorescence values reflect protein concentration, up to a proportionality factor (intensity of a fluorophore). The fractional measurement noise in estimating relative concentrations can be estimated as the standard deviation of pixel intensity values within a nucleus on the projected map. In their respective regions of expression, this standard deviation of Hb, Kr and Eve pixel intensity constituted $\approx 1\%$ of the expression value and was therefore negligible compared to the expression noise observed across nuclei (Fig. 4B). To avoid signal distortion artifacts observed at the edges of the imaged portion of the embryo due to tissue curvature and compression, all analysis was restricted to nuclei located in the low-distortion region selected manually along the imaged embryo center line, typically 20-25 nuclei wide (Fig.~\ref{fig:embryo8}).

\subsection{Estimating expression noise (Fig. 4B)\label{app:noiseCalcDetails}}

Expression noise is defined as:
$$
c_\noise = c_\recorded - c_\expected,
$$
where $c_\recorded$ is the recorded fluorescent intensity (of Hb, Kr or Eve), and $c_\expected$ is the expected value at that location. Measuring noise therefore requires a method for constructing $c_\expected$. We use a method that we call ``haltere-shaped filtering''. To introduce and motivate this method, we begin by discuss two simpler alternatives and their limitations: binning by AP coordinate and neighbor averaging.

\begin{figure}[b!]
\centering
\includegraphics[width=0.95\linewidth]{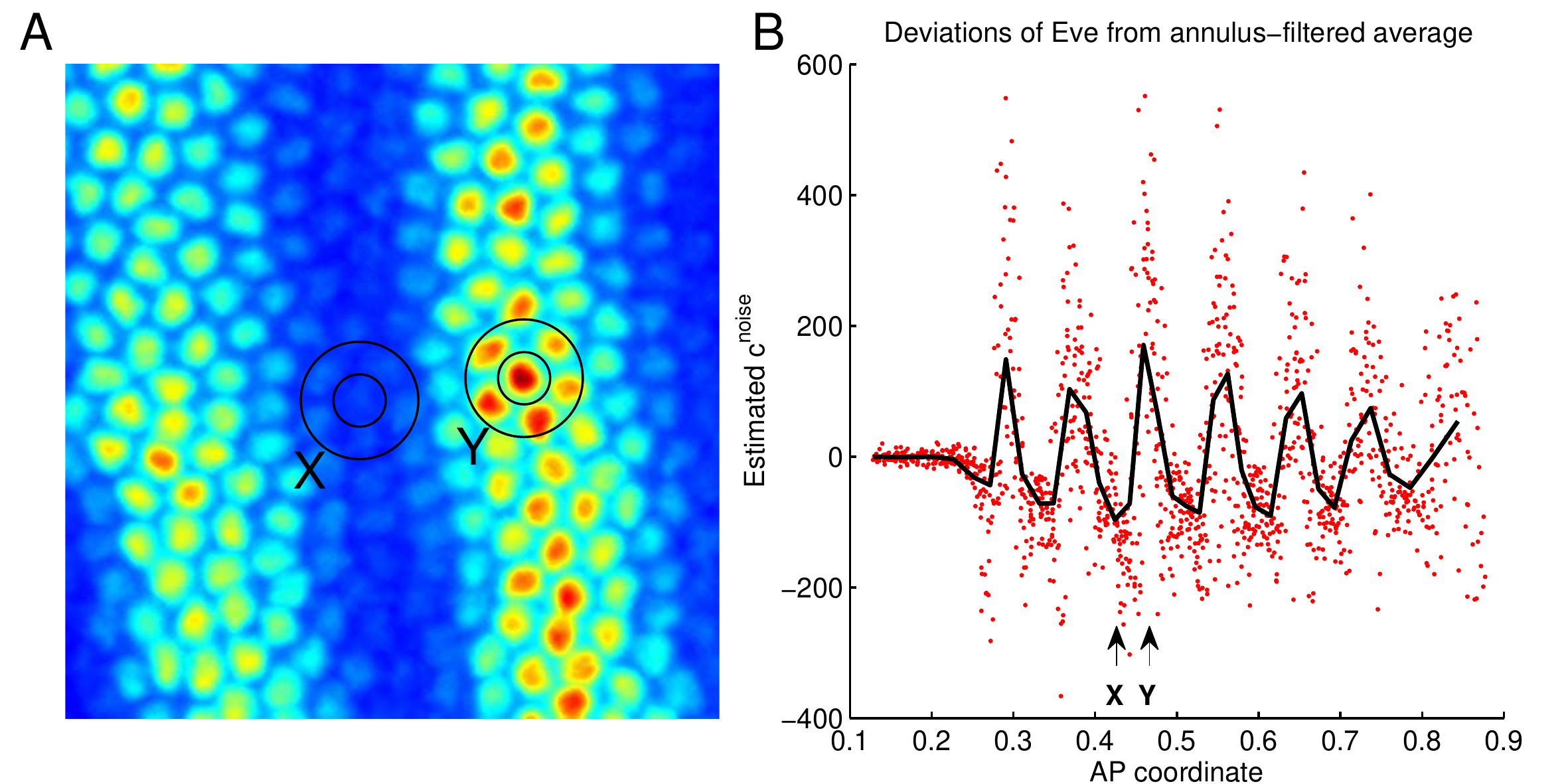}
\caption{\small{The simple neighbor-averaging method will underestimate $c_\expected$ in the regions where the profile is concave, e.g.\ at the peaks of Eve stripes (nucleus X), and overestimate $c_\expected$ where the profile is convex, e.g.\ in the Eve troughs (nucleus Y). \textbf{A}: Eve stripes 2 and 3. Nuclei X and Y marked by smaller circles; the large circles encompass the neighbors over which averaging is performed. \textbf{B}: $c_\noise$ as estimated using the neighbor-averaging method, shown as a function of AP coordinate. Black line: window average of $c_\noise$ over 50 consecutive nuclei. This average should be close to zero for an unbiased estimate, but exhibits a clear correlation with the Eve profile shape.}\label{fig:annulus}}
\end{figure}

\subsubsection{Binning by AP coordinate}
Since gap genes expression is often said to be a function of the location along the antero-posterior (AP) axis, one approach could be to define $c_\expected$ as the average expression level in all nuclei with a similar AP coordinate. This approach, however, would yield strongly biased results due to the curvature of gene expression domains (Fig.~\ref{fig:embryo8}).

\subsubsection{Neighbor averaging}

A better approach is to construct $c_\expected$ for each nucleus based on the expression levels observed in neighboring nuclei. Since expression profiles are relatively smooth functions of location, the average of expression levels in nuclei that are immediate neighbors of nucleus $i$ provides a reasonable expectation for $c_i$. Despite being a significant improvement over the naive AP-based method, however, the simple averaging over neighbors provides an unbiased estimate only in regions where the profile shape is well approximated by a linear dependence. In all other cases this estimate will have a bias proportional to the convexity (second derivative) of the mean profile shape. This is particularly clear for the sharply varying profile of Eve (Fig.~\ref{fig:annulus}A). This bias can lead to a dangerous artifact, whereby sharply varying profiles would appear to be more noisy, which would be unacceptable for our analysis of the Hb-Kr-Eve system. Fig.~\ref{fig:annulus}B shows the inferred $c_\noise$ as a function of AP axis coordinate. The severity of the bias of the neighbor-averaging method of estimating $c_\expected$ can be measured by the clearly observed correlation between $c_\noise$ and the average profile shape of Eve (i.e.\ $c_\recorded$).

\subsubsection{Haltere-shaped filtering}

\begin{figure}[t!]
\centering
\includegraphics[width=0.95\linewidth]{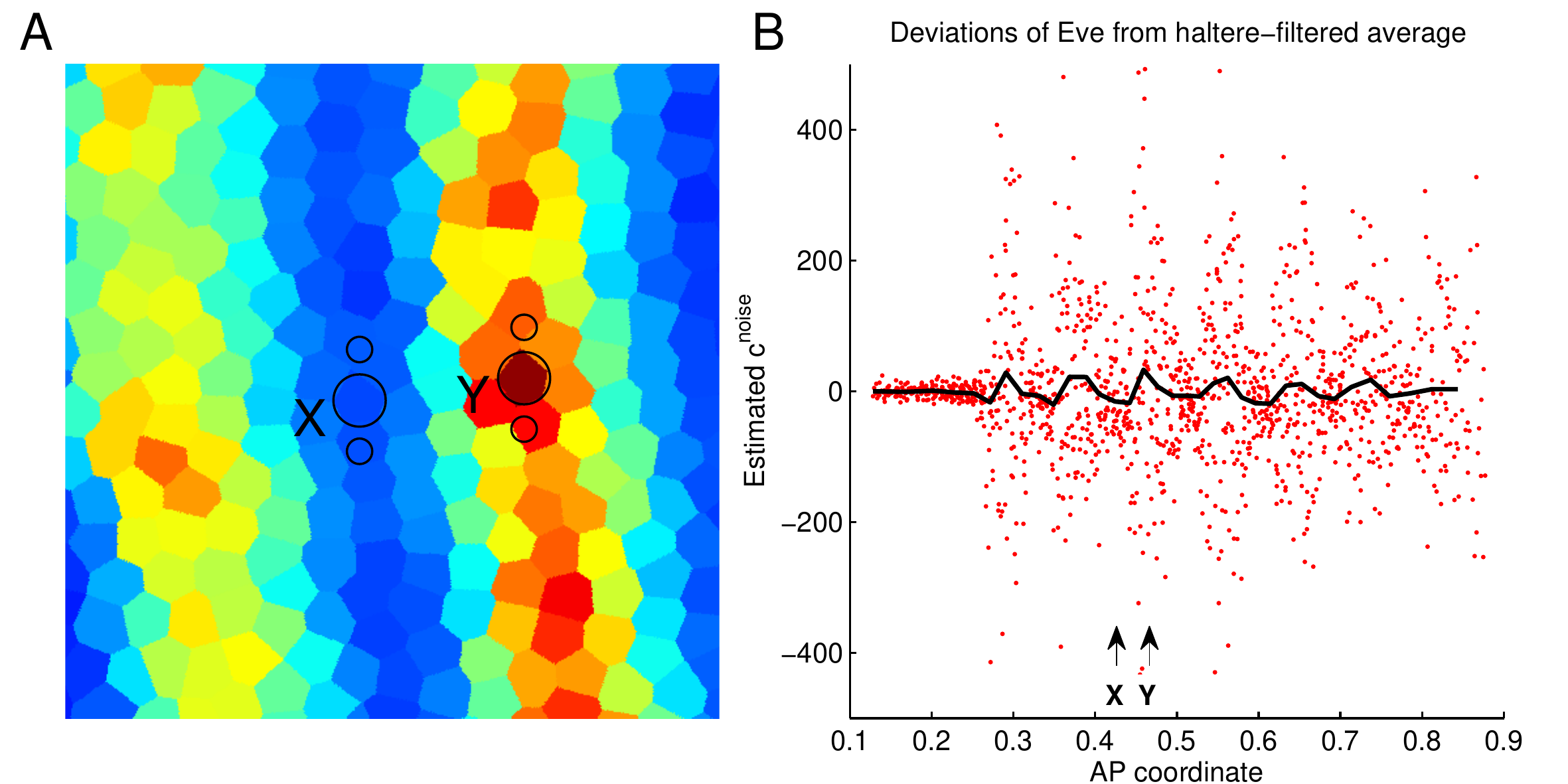}
\caption{\small{\textbf{A}: ``Eve map'' of the region depicted in Fig.~\ref{fig:annulus}A, constructed as described in the text. X and Y label the same nuclei as in Fig.~\ref{fig:annulus}A; the larger circle marks their location. The smaller circles depict the haltere-shaped filter: $c_\expected$ is constructed as the average pixel value over this area around each nucleus. \textbf{B}: Inferred $c_\noise$ shown as a function of AP coordinate. The performance of the haltere-filtering method shows marked improvement compared to annulus filtering (Fig.~\ref{fig:annulus}B), as indicated by the greatly reduced fluctuations of the window-averaged $c_\noise$ (in black). The fact that the magnitude of $c_\noise$ increases in regions of greater expression is normal: larger expression means larger absolute noise.}\label{fig:haltere}}
\end{figure}

We now describe the procedure we used to construct $c_\expected$ for our analysis. We begin by creating an ``expression map'' whereby in the projected image such as depicted in Fig.~\ref{fig:embryo8} the value of every pixel is replaced by the expression level $c_\recorded$ recorded in the nucleus closest to that pixel. The image is then filtered using a haltere-shaped filter depicted in Fig.~\ref{fig:haltere}A, and pixel values at each nucleus after filtering define the values of $c_\expected$.

This method combines the better qualities of the two approaches discussed above. On a perfectly regular hexagonal lattice, this would be equivalent to the neighbor-averaging method using only the immediate dorsal and ventral neighbors, but the specific procedure we described naturally deal with lattice imperfections. In fact, $c_\noise$ in Fig.~\ref{fig:annulus}B was constructed using this exact procedure, but using an annulus-shaped filter depicted in Fig.~\ref{fig:annulus}A. Since the gradient of expression profiles is predominantly aligned with the AP axis, using a haltere-shaped filter greatly reduces any introduced bias (Fig.~\ref{fig:haltere}B).

One might expect that for even higher accuracy, the orientation of the haltere filter could be set not by perpendicularity to the imaginary AP axis, but by the isolines of the actual expression profile after sufficiently strong smoothing. However, in practice such an approach is functionally less robust due to the number of tunable parameters, and we empirically found the fixed-angle haltere filtering to result in the lowest bias as measured by the correlation of average $c_\noise$ in a region and the average $c_\recorded$ in that same region.

\subsection{Idealized profiles (Fig. 4C)}
The expression profiles of long body axis patterning genes in \emph{Drosophila} form a pattern that, to a good approximation, can be considered one-dimensional. However, as discussed above, due to the curvature of expression profiles, $x_\AP$ is not the variable that best captures the variance. To estimate positional information in a gene expression pattern using data from single embryos, we therefore use the measured expression pattern shape and noise to construct what we call ``idealized profiles''. First, we plot the recorded expression values $c_\recorded$ as a function of $x_\AP$ and construct a smooth spline fit that captures the mean profile shape; we denote the result $\mu(x_\AP)$. Next, the same procedure is applied to expression noise, estimated as described above: the smooth spline fit to $c_\noise^2$ as a function of $x_\AP$ describes how the experimentally observed expression noise varies along the AP axis; we denote this root-mean-square deviation function $e(x_\AP)$. An expression pattern with mean $\mu(x_\AP)$ and independent Gaussian noise of magnitude $e(x_\AP)$ constitutes the ``idealized profile'' of a given patterning cue (see Fig.~4C).

Note that when calculating average noise magnitude for a given AP coordinate, expression noise is calculated as described in the previous section, i.e.\ \emph{prior} to binning by AP. The result is the average of expression noise measured locally for all nuclei at a similar AP location --- as opposed to the variance of expression among all nuclei at the same $x_\AP$; the latter, as we described, suffers from artifacts. The procedure we described effectively straightens out expression stripes: the resulting profile has the same mean and noise magnitude as observed experimentally, but is, by construction, a function of a single variable. This approach contrasts with the procedure of~\cite{Dubuis:2013} where embryos were imaged in cross-section and only dorsal or ventral ``expression profiles'' were used, i.e. expression levels were recorded along a \emph{particular} AP line (from multiple embryos). Here, we use \emph{all} nuclei observed on a slightly flattened surface of a single embryo, and the variation of expression profile shape with the dorsal-ventral coordinate becomes a major factor.

\subsection{Computing information content (Fig. 4D)}
By definition, the information content (or the mutual information) $I(c,x)$ of a profile $c(x)$ is the average reduction of uncertainty of $c$ after $x$ becomes known:
$$
I(x,c)=S(c) - \langle S(c|x) \rangle_x.
$$
Here the first term is the entropy of the full distribution of $c$, which we denote $P_c$, and $S(c|x)$ is the entropy of the conditional distribution $P(c|x)$. We write:
$$P_c(c)=\int\! p(c|x) P_x(x) \,dx = \frac 1{x_\mathrm{min}-x_\mathrm{max}}\int\! p(c|x)\,dx,$$
because the position $x$ is uniformly distributed between $x_\mathrm{min}$ and $x_\mathrm{max}$ (in our case, $x_{\AP\,\mathrm{min}}=0.37$ and $x_{\AP\,\mathrm{max}}=0.47$).

These formulas express the information content of a one-dimensional profile entirely in terms of the conditional probability function $p(c|x)$. For the idealized profile, at a given AP location $x_0$, the conditional distribution $p(c|x_0)$ is Gaussian with mean $\mu(x_0)$ and width $e(x_0)$; in particular, the entropy of $p(c|x_0)$ is known analytically. Therefore, we compute $I(x,c)$ by numerically performing the integral. We validated our code by computing information content of simple profiles for which the information content can also be calculated analytically.


\begin{thebibliography}{99}
%
\bibitem{johnson_70}
Johnson HA (1970)
Information theory in biology after 18 years. {\em Science}, 168:1545--1550.
%
\bibitem{waltermann_11}
Waltermann C, Klipp E (2011)
Information theory based approaches to cellular signaling. {\em Biochim Biophys Acta}, 1810(10):924--32.
%
\bibitem{bowsher_14}
Bowsher CG, Swain PS (2014)
Environmental sensing, information transfer, and cellular decision-making. {\em Curr Opin Biotechnol}, 28:149--55.
%
\bibitem{levchenko_14}
Levchenko A, Nemenman I (2014)
Cellular noise and information transmission. {\em Curr Opin Biotechnol}, 28:156--164.
%
\bibitem{tkacikBialek_14}
Tkacik G, Bialek W (2014)
Information processing in living systems. arXiv:1412.8752.
%
\bibitem{cheong_11}
Cheong R, Rhee A, Wang CJ, Nemenman I, Levchenko A (2011)
Information transduction capacity of noisy biochemical signaling networks. {\em Science} 334(6054):354--8.
%
\bibitem{selimkhanov_14}
Selimkhanov J, Taylor B, Yao J, Pilko A, Albeck J, Hoffmann A, Tsimring L, Wollman R (2014)
Systems biology. Accurate information transmission through dynamic biochemical signaling networks.
{\em Science} 346(6215):1370--3.
%
\bibitem{dubuis_13}
Dubuis JO, Tkacik G, Wieschaus EF, Gregor T, Bialek W (2013)
Positional information, in bits. {\em Proc Natl Acad Sci U~S~A} 110:16301--8.
%
\bibitem{hironaka_2012}
Hironaka K, Morishita Y (2012)
Encoding and decoding of positional information in morphogen-dependent patterning. {\em Curr Opin Genet Dev} 22:553--561.
%
\bibitem{lander_2013}
Lander A (2013)
How cells know where they are. {\em Science} 339:923--927.
%
\bibitem{rogers_2011}
Rogers KW, Schier AF (2011)
Morphogen gradients: from generation to interpretation. {\em Annu Rev Cell Dev Biol} 27:377--407.
%
\bibitem{nahmad_2011}
Nahmad M, Lander AD (2011)
Spatiotemporal mechanisms of morphogen gradient interpretation. {\em Curr Opin Genet Dev} 21:726--731.
%
\bibitem{wartlick_2009}
Wartlick O, Kicheva A, Gonzalez-Gaitan M (2009)
Morphogen gradient formation. {\em Cold Spring Harb Persp Biol} 1:a001255.
%
\bibitem{muller_2013}
Muller P, Rogers KW, Yu SR, Brand M, Schier AF (2013)
Morphogen transport. {\em Development} 140:1621--1638.
%
\bibitem{wolpert_1969}
Wolpert L (1969)
Positional information and the spatial pattern of cellular differentiation. {\em J Theor Biol} 25:1--47.
%
\bibitem{shannon_48}
Shannon, CE (1948)
A mathematical theory of communication. {\em Bell Systems Technical J} 27:379--423, 623--656.
%
\bibitem{gergen_86}
Gergen JP, Coulter D, Wieschaus EF (1986)
Segmental pattern and blastoderm cell identities. \emph{Gametogenesis and the Early Embryo}, ed J. Gall (Alan R. Liss, New York) pp. 195--220.
%
\bibitem{gregor_2007}
Gregor T, Tank DW, Wieschaus EF, Bialek W (2007)
Probing the limits to positional information. {\em Cell} 130:153--164.
%
\bibitem{munsky_2012}
Munsky B, Neuert G, van Oudenaarden A (2012)
Using gene expression noise to understand gene regulation. {\em Science} 336:183--187.
%
\bibitem{sanchez_2013}
Sanchez A, Golding I (2013)
Genetic determinants and cellular constraints in noisy gene expression. {\em Science} 342:1188--1193.
%
\bibitem{pera_14}
Pera EM, Acosta H, Gouignard N, Climent M, Arregi I (2014)
Active signals, gradient formation and regional specificity in neural induction.{\em Exp Cell Res} 321:25--31.
%
\bibitem{lumsden_96}
Lumsden A, Krumlauf R (1996)
Patterning the Vertebrate Neuraxis. {\em Science} 274:1109--1115.
%
\bibitem{raible_04}
Raible F, Brand M (2004)
{\em Divide et Impera} -- the midbrain--hindbrain boundary and its organizer. {\em Trends Neurosci} 27:727--734.
%
\bibitem{patthey_14}
Patthey C, Gunhaga L (2014)
Signaling pathways regulating ectodermal cell fate choices. {\em Exp Cell Res} 321:11--16.
%
\bibitem{saga_12}
Saga Y (2012)
The mechanism of somite formation in mice. {\em Curr Opin Genet Dev} 22:331--338.
%
\bibitem{watanabe_10}
Watanabe T, Takahashi Y (2010)
Tissue morphogenesis coupled with cell shape changes. {\em Curr Opin Genet Dev} 20:443--447.
%
\bibitem{little_06}
Little SC, Mullins MC (2006)
Extracellular modulation of BMP activity in patterning the dorsoventral axis. {\em Birth Defects Res C Embryo Today} 78:224--242.
%
\bibitem{rushlow_12}
Rushlow CA, Shvartsman SY (2012)
Temporal dynamics, spatial range, and transcriptional interpretation of the Dorsal morphogen gradient. {\em Curr Opin Genet Dev} 22:542--546.
%
\bibitem{driever_88}
Driever W, Nusslein--Volhard C (1988)
A gradient of bicoid protein in Drosophila embryos. {\em Cell} 54:83--93.
%
\bibitem{kornberg_93}
Kornberg TB, Tabata T (1993)
Segmentation of the Drosophila embryo. {\em Curr Opin Genet Dev} 3:585--593.
%
\bibitem{tkacik_15}
Sokolowski TR, Tkacik G (2015)
Optimizing information flow in small genetic networks. IV. Spatial coupling. arXiv:1501.04015.
%
\bibitem{tostevin_09}
Tostevin F, ten Wolde PR (2009)
Mutual information between input and output trajectories of biochemical networks. {\em PRL} 102:218101.
%
\bibitem{tkacik_14}
Tkacik G, Dubuis JO, Petkova MD, Gregor T (2015)
Positional information, positional error, and read-out precision in morphogenesis: a mathematical framework.
{\em Genetics} 199(1): 39--59.
%
\bibitem{petkova_14}
Petkova MD, Little SC, Liu F, Gregor T (2014)
Maternal origins of developmental reproducibility. {\em Current Biology} 24:1283–1288.
%
\bibitem{little_13}
Little SC, Tikhonov M, Gregor T (2013)
Precise developmental gene expression arises from globally stochastic transcriptional activity. {\em Cell} 154:789--800.
%
\bibitem{tenWolde_09}
Erdmann T, Howard M, ten Wolde PR (2009)
Role of spatial averaging in the precision of gene expression patterns. {\em Phys Rev Lett} 103:258101.
%
\bibitem{gilbert}
Gilbert SF (2013)
{\em Developmental Biology.} (Sinauer Associates, Inc.), 10th edition.
%
\bibitem{tierIntegrate1}
Li XY et al. (2008)
Transcription factors bind thousands of active and inactive regions in the {\em Drosophila} blastoderm. {\em PLoS Biol.} 6(2):e27.
%
\bibitem{tierIntegrate2}
MacArthur S et al. (2009)
Developmental roles of 21 Drosophila transcription factors are determined by quantitative differences in binding to an overlapping set of thousands of genomic regions. {\em Genome Biol.} 10(7):R80.
%
\bibitem{noteJointAcc}
For multiple profiles $\{c^{(1)},c^{(2)},\dots\}$, we define accessible information as the joint information content in the set of morphogen profiles, independently corrupted with noise of magnitude $\eta_0$ (compare with Eq.~\eqref{eq:defAcc}):
$$
I_\acc\big(\{c^{(1)},c^{(2)},\dots\}\big)\equiv I_\raw\big(\{c^{(1)}+\eta^{(1)},c^{(2)}+\eta^{(2)},\dots\}\big).
$$
%
\bibitem{tkacik_08}
Tkacik G, Gregor T, Bialek W (2008)
The role of input noise in transcriptional regulation. {\em PLoS ONE} 3(7): e2774.
%
\bibitem{krivega_2012}
Krivega I, Dean A (2012)
Enhancer and promoter interactions -- long distance calls. {\em Curr Opin Genet Dev} 22:79--85.
%
\bibitem{kwak_2013}
Kwak H, Lis JT (2013)
Control of transcriptional elongation. {\em Annu Rev Genet} 47:483--508.
%
\bibitem{maheshri_2007}
Maheshri N, O'Shea EK (2007)
Living with noisy genes: how cells function reliably with inherent variability in gene expression. {\em Annu Rev Biophys Biomol Struct} 36:413--434.
%
\bibitem{kraut_91}
Kraut R, Levine M (1991)
Spatial regulation of the gap gene giant during Drosophila development. {\em Development} 111:601--609.
%
\bibitem{small_96}
Small S, Blair A, Levine M (1996)
Regulation of two pair-rule stripes by a single enhancer in the Drosophila embryo. {\em Dev Biol} 175:314-324.
%
\bibitem{bowsher_13}
Bowsher CG, Voliotis M, Swain PS (2013)
The fidelity of dynamic signaling by noisy biomolecular networks. {\em PLoS Comput Biol}, 9(3):e1002965.
%
\bibitem{deronde_14}
de Ronde W, ten Wolde PR (2014)
Multiplexing oscillatory biochemical signals. {\em Phys Biol}, 11(2):026004.
%
\bibitem{optimFlow1}
Tkacik G, Walczak AM, Bialek W (2009)
Optimizing information flow in small genetic networks. {\em Phys Rev E} 80:031920.
%
\bibitem{optimFlow2}
Walczak AM, Tkacik G, Bialek W (2010)
Optimizing information flow in small genetic networks II. Feed-forward interactions. {\em Phys Rev E} 81:041905.
%
\bibitem{optimFlow3}
Tkacik G, Walczak AM, Bialek W (2012)
Optimizing information flow in small genetic networks III. A self-interacting gene. {\em Phys Rev E} 85:041903.
%
\bibitem{macKay}
MacKay DJC (2012)
\textit {Information theory, inference and learning algorithms} (Cambridge University Press), Fig.~47.17, p.~568.
\end{thebibliography}

\begin{thebibliography}{99}
\bibitem{Dubuis:2013}
Dubuis JO, Samanta R, Gregor T (2013)
Accurate measurements of dynamics and reproducibility in small genetic networks. {\em Mol Syst Biol} 9:639.
\bibitem{DubuisTkacik:2013}
Dubuis JO, Tkacik G, Wieschaus EF, Gregor T, Bialek W (2013)
Positional information, in bits. {\em Proc Natl Acad Sci U~S~A} 110, 16301-16308.
\end{thebibliography}
\end{document}